\documentclass[12pt]{iopart}
\usepackage{graphicx}
\usepackage{amssymb}
\begin{document}

\title[Exactly solved spin-$1$ Ising--Heisenberg diamond chain]{Magnetization plateaus of an exactly solvable spin-$1$ Ising--Heisenberg diamond chain}

\author{N S Ananikian$^{1,2}$, J Stre\v{c}ka$^3$, and V V Hovhannisyan$^1$}

\address{$^1$ A.I. Alikhanyan National Science Laboratory, 0036 Yerevan, Armenia}
\address{$^2$ Applied Mathematics Research Centre, Coventry University, Coventry, CV1 5FB, England, UK}
\address{$^3$ Institute of Physics, Faculty of Science, P. J. \v{S}af\'{a}rik University, Park Angelinum 9, 040 01, Ko\v{s}ice, Slovak Republic}
\eads{\mailto{jozef.strecka@upjs.sk}, \mailto{vahmer@rambler.ru}}

\begin{abstract}
The spin-1 Ising--Heisenberg diamond chain in a magnetic field is exactly solved by a rigorous treatment based on the transfer-matrix method. An exact ground-state phase diagram includes in total three unconventional quantum ground states with a quantum entanglement of the decorating spin-1 Heisenberg dimers apart from two ground states with a classical spin arrangement. It is evidenced that all three values of the magnetization allowed for the spin-1 diamond chain without translationally broken symmetry by the Oshikawa-Yamanaka-Affleck criterion  can become evident in an outstanding stepwise magnetization curve with three intermediate plateaus at zero, one-third, and two-thirds of the saturation magnetization.
\end{abstract}

\pacs{05.50.+q, 75.10.Hk, 75.10.Jm, 75.10.Pq, 75.40.Cx}
\submitto{\JPCM}


\section{Introduction}

Over the last few years, the spin-1/2 Ising--Heisenberg diamond
chain has become a subject of intensive theoretical studies by
virtue of its remarkable magnetic properties, which arise out from a
mutual interplay between a geometric spin frustration and quantum
fluctuations
\cite{can06,ana12,roj12,bel13,ana13,ana14,dua14,val08,lis11,gal13,gal14,lis14}.
Among the most notable features of this exactly solvable spin chain
one could mention an existence of one-third magnetization plateau in
a low-temperature magnetization curve, spectacular temperature
dependences of susceptibility and specific heat \cite{can06},
outstanding temperature and magnetic-field dependences of
entanglement measures \cite{ana12,roj12}, highly non-monotonous
thermal variations of correlation functions \cite{bel13}, a striking
plateau of Lyapunov exponent \cite{ana13}, a  remarkable
distribution of partition function zeros \cite{ana14} and an
enhanced cooling rate during the adiabatic demagnetization
\cite{dua14}. It turns out, moreover, that generalized versions of
the spin-1/2 Ising--Heisenberg diamond chain accounting for the
asymmetry \cite{val08,lis11}, four-spin coupling \cite{gal13,gal14}
and/or second-neighbour interaction \cite{lis14} might exhibit even
a more diverse magnetic behaviour. The existance of the
magnetization plateaus and multiple peak structure of the specific
heat have been also detected on a distorted Ising-Hubbard diamond
chain \cite{ana14a}. In spite of a certain over-simplification, the
asymmetric version of the spin-1/2 Ising--Heisenberg diamond chain
accounting also for the second-neighbour coupling \cite{lis14}
quantitatively reproduces several experimentally observed magnetic
features of the natural mineral azurite, which is regarded to date
as the most famous representative of the spin-1/2 diamond chain
\cite{kik03,kik04,kik05a,kik05b}.

It has been argued by Oshikawa, Yamanaka, and Affleck (OYA)
\cite{osh97,aff98} that fractional values of the magnetization at
intermediate magnetization plateaus of a quantum spin chain must
necessarily satisfy the condition $p(S_u - m_u) = \mathbb{Z}$, where
$p$ is a period of the ground state, $S_u$ and $m_u$ denote the
total spin and total magnetization per elementary unit,
respectively. In accordance with the OYA rule, the spin-1/2
Ising--Heisenberg diamond chain can exhibit an intermediate plateau
just at one-third of the saturation magnetization on assumption that
a translational symmetry of the spin-1/2 diamond chain is not broken
(i.e. for the period of ground state $p=1$). However, the extended
versions of the spin-1/2 Ising--Heisenberg diamond chain accounting
for the asymmetry, four-spin coupling and/or second-neighbour
interaction can quite naturally cause a doubling of the period of
ground state, which consequently allows according to the OYA
criterion additional magnetization plateaus at zero and two-thirds
of the saturation magnetization in addition to the one-third
magnetization plateau. The similar situation takes place for the
spin-1/2 Heisenberg chain with hexamer modulation of exchange where
the existence of two new magnetization plateaus at $1/3$ and $2/3$
have been observed \cite{gia12}. It has been actually evidenced that
the spatial asymmetry along diamond sides can be responsible for a
presence of the intermediate plateau at zero magnetization
\cite{lis11}, while the four-spin coupling and second-neighbour
interaction can cause emergence of both intermediate plateaus at
zero and two-thirds of the full magnetization
\cite{gal13,gal14,lis14}.

The OYA criterion would suggest that an intriguing magnetization curve with a greater number of intermediate magnetization plateaus can be alternatively obtained by increasing the total spin per unit cell. From this perspective, it is worthwhile to consider a symmetric version of the spin-$1$ Ising--Heisenberg diamond chain, which may capture intermediate plateaus at zero, one-third and/or two-thirds of the saturation magnetization even if a translational symmetry is restored (i.e. the period of ground state is not doubled due to the asymmetry, multispin and/or further-neighbour interactions). The main goal of the present article is to verify whether or not all intermediate magnetization plateaus admissible by the OYA rule can be indeed detected in a magnetization process of the symmetric spin-$1$ Ising--Heisenberg diamond chain.

The paper is organized as follows. In Sec. \ref{model} we will introduce the investigated spin-chain model and briefly describe basic steps of our rigorous calculation. The most interesting results for the ground state, magnetization process and susceptibility are discussed in detail in Sec. \ref{result}. Finally, the conclusions and future outlooks are briefly mentioned in Sec. \ref{conclusion}.

\section{The model and its exact solution}
\label{model}

We consider the spin-$1$ Ising-Heisenberg model on a diamond chain in a presence of the external magnetic field. The primitive unit cell of a diamond chain consists of two Heisenberg spins $S_{a,i}$ and $S_{b,i}$ that interact symmetrically via Ising-type interaction with two nearest-neighbour Ising spins
$\mu_{i}$ and $\mu_{i+1}$ (see Fig.~1). The total Hamiltonian of the model under investigation may be represented as a sum over block Hamiltonians $\mathcal{H}=\sum_{i=1}^{N}\mathcal{H}_{i}$, where
\begin{eqnarray}
\mathcal{H}_{i} &=&
J[\Delta(S_{a,i}^{x}S_{b,i}^{x}+S_{a,i}^{y}S_{b,i}^{y})+
S_{a,i}^{z}S_{b,i}^{z}]+J_{1}\left(S_{a,i}^{z}+S_{b,i}^{z}\right)\left(\mu_{i}+\mu_{i+1}\right) \nonumber \\
&-&H_{H}\left(S_{a,i}^{z}+S_{b,i}^{z}\right) -H_{I}\frac{\mu_{i}+\mu_{i+1}}{2}.
\label{1.1}
\end{eqnarray}
In above, $S_{a,i}^{\alpha}$ and $S_{b,i}^{\alpha}$ ($\alpha=x,y,z$) denote spatial components of the spin-$1$ operators, $\mu_{i} = \pm1,0$ stands for the Ising spin, $J$ labels the XXZ interaction between the nearest-neighbour Heisenberg spins, $\Delta$ is a spatial anisotropy in this interaction, $J_1$ is the Ising interaction between the nearest-neighbour Ising and Heisenberg spins and finally, the last two terms determine Zeeman's energy of the Heisenberg and Ising spins in a longitudinal magnetic field.

\begin{figure}
\includegraphics[scale=0.7]{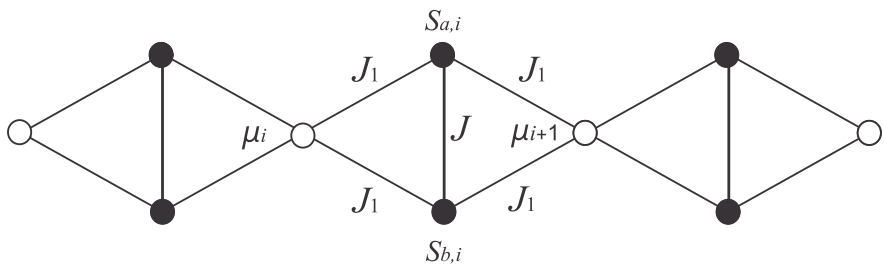}
\caption{A schematic representation of the spin-1 diamond chain. The Ising ($\mu_{i}$, $\mu_{i+1}$) and Heisenberg ($S_{a,i}$, $S_{b,i}$) spins belonging to the $i$-th block of the diamond chain are marked.}
\label{fig:diamond}
\end{figure}

The important part of our further calculations is based on the commutation relation between different block Hamiltonians $[\mathcal{H}_i,\mathcal{H}_j]=0$, which will allow us to partially factorize the partition function of the model and represent it as a product over block partition functions
\begin{eqnarray}
Z=\sum_{\mu_{i}}\prod^{N}_{i=1}\mbox{Tr}_{i}\rme^{-\beta \mathcal{H}_{i}},
\label{1.2}
\end{eqnarray}
where $\beta=(k_{B}T)^{-1}$, $k_{B}$ is Boltzmann's constant, $T$ is the absolute temperature, $\sum_{\mu_{i}}$ marks a summation over spin states of all Ising spins and $\mbox{Tr}_{i}$ means a trace over the spin degrees of freedom of two Heisenberg spins from the $i$-th block. After a straightforward diagonalization of the cell Hamiltonian (\ref{1.1}) of the spin-1 quantum Heisenberg dimer one obtains the following expressions for the respective eigenvalues
\begin{eqnarray}
\mathcal{E}_{1}(\mu_{i},\mu_{i+1})= & -J-\frac{H_I}{2}(\mu_i+\mu_{i+1}),\nonumber \\
\mathcal{E}_{2,3}(\mu_{i},\mu_{i+1})= & \pm J\Delta - J_1(\mu_i+\mu_{i+1}) -\frac{H_I}{2}(\mu_i+\mu_{i+1})+H_H,\nonumber \\
\mathcal{E}_{4,5}(\mu_{i},\mu_{i+1})= & \pm J\Delta + J_1(\mu_i+\mu_{i+1}) -\frac{H_I}{2}(\mu_i+\mu_{i+1})-H_H,\nonumber \\
\mathcal{E}_{6,7}(\mu_{i},\mu_{i+1})= & -\frac{J}{2}(1 \pm \delta) - \frac{H_I}{2}(\mu_i+\mu_{i+1}) ,\nonumber \\
\mathcal{E}_{8,9}(\mu_{i},\mu_{i+1})= &  J \pm 2J_1(\mu_i+\mu_{i+1})
-\frac{H_I}{2}(\mu_i+\mu_{i+1}) \mp 2H_H,
\label{1.4}
\end{eqnarray}
where $\delta=\sqrt{1+8\Delta^2}$. Now, one may simply perform a trace over the spin degrees of freedom of the spin-1 Heisenberg dimers on the right-hand-side of Eq.~(\ref{1.2}) and the partition function can be consequently rewritten into the following form
\begin{eqnarray}
Z=\sum_{\mu_{i}}\prod^{N}_{i=1}T_{\mu_{i},\mu_{i+1}} = \mbox{Tr} \,T^{N},
\label{1.5}
\end{eqnarray}
where the expression $T_{\mu_{i},\mu_{i+1}}$ can be viewed the standard $3 \times 3$ transfer matrix
\begin{eqnarray}
T_{\mu_{i},\mu_{i+1}} =  \left(
\begin{array}{cccc}
T_{1,1}  & T_{1,0} & T_{1,-1}  \\\
T_{0,1} & T_{0,0} & T_{0,-1} \\\
T_{-1,1}  & T_{-1,0} & T_{-1,-1}
\end{array} \right).
\label{1.7}
\end{eqnarray}
Here, $\pm1$ and 0 denote three spin states of the Ising spins $\mu_i=\pm1$ and $0$, whereas the transfer matrix (\ref{1.7}) has precisely the same form as the transfer matrix of the generalized spin-1 Blume-Emery-Griffiths chain diagonalized in Refs.~\cite{kri74,kri75,lis13}. Of course, individual elements of the transfer matrix (\ref{1.7}) are defined through the formula
\begin{eqnarray}
T_{\mu_{i},\mu_{i+1}}=\mbox{Tr}_{i}\rme^{-\beta \mathcal{H}_{i}}=\sum_{n=1}^{9}\rme^{-\beta\mathcal{E}_{n}(\mu_{i},\mu_{i+1})},
\label{1.6}
\end{eqnarray}
which includes the set of eigenvalues (\ref{1.4}) for the spin-1 quantum Heisenberg dimer. With regard to Eq.~(\ref{1.5}), the partition function of
the spin-1 Ising-Heisenberg diamond chain can be expressed through three eigenvalues of the transfer matrix (\ref{1.7}) explicitly given in Refs.~\cite{kri74,kri75,lis13}
\begin{eqnarray}
Z=\lambda_{1}^{N}+\lambda_{2}^{N}+\lambda_{3}^{N}.
\label{1.8}
\end{eqnarray}
Next, let us denote the largest transfer-matrix eigenvalue $\lambda= \mbox{max} \{ \lambda_1, \lambda_2,\lambda_3 \}$, then, the contribution of two smaller transfer-matrix eigenvalues to the partition function may be completely neglected in the thermodynamic limit $N \rightarrow \infty$
\begin{eqnarray}
Z \simeq \lambda^{N}.
\label{1.9}
\end{eqnarray}
The free energy per elementary diamond cell can be obtained from the largest eigenvalue of the transfer matrix (\ref{1.6}) according to the formula
\begin{eqnarray}
f=-\lim_{N \to \infty} \frac{1}{\beta N} \ln Z = -\frac{1}{\beta}\ln\lambda.
\label{1.10}
\end{eqnarray}
Other thermodynamic quantities of the investigated spin system (such as magnetization, magnetic susceptibility, specific heat) can be obtained in terms of
the free energy and its derivatives. The sublattice magnetization of Ising and Heisenberg spins readily follows from the formulas
\begin{eqnarray}
m_I=\frac{1}{Z}\sum_{\mu_{i}}\mbox{Tr}_{i}\left(\mu_i \rme^{-\beta
\mathcal{H}}\right)=-\left(\frac{\partial f}{\partial H_I}
\right)_{T, H_H},
\nonumber \\
m_H=\frac{1}{2Z}\sum_{\mu_{i}}\mbox{Tr}_{i}\left((S_{a,i}^z+S_{b,i}^z)
\rme^{-\beta \mathcal{H}}\right)=-\left(\frac{\partial f}{\partial H_H}
\right)_{T, H_I}. \label{1.11}
\end{eqnarray}
In this regard, the total magnetization of the spin-1 Ising-Heisenberg chain per one spin can be computed from the relation
\begin{eqnarray}
m=\frac{1}{3Z}\sum_{\mu_{i}}\mbox{Tr}_{i}\left((\mu_i^z+S_{a,i}^z+S_{b,i}^z)
\rme^{-\beta \mathcal{H}}\right)=\frac{1}{3}m_I+\frac{2}{3}m_H.
 \label{1.12}
\end{eqnarray}

\section{Results and discussion}
\label{result}

In this section, we will focus our attention to a rigorous analysis of the ground state and the most important magnetic properties of the spin-1 Ising-Heisenberg diamond chain with the antiferromagnetic exchange interactions $J>0$ and $J_1>0$. Hereafter, we will consider for simplicity the uniform external magnetic field acting on the Ising and Heisenberg spins $H_{H}=H_{I}\equiv H$, whereas a strength of the Ising coupling $J_1$ will be used for introducing a set of dimensionless parameters $\alpha= J/J_1$, $h=H/J_1$ $t=k_B T/J_1$ measuring a relative strength of both exchange constants, magnetic field and temperature, respectively. It is noteworthy that the antiferromagnetic Heisenberg interaction brings a geometric spin frustration into play, so the newly defined interaction ratio $\alpha$ can be also viewed as the frustration parameter.

\begin{figure}[t]
\begin{center}
\begin{tabular}{ccc}
{\small (a)}& {\small (b)}\\
\includegraphics[width=5.1cm]{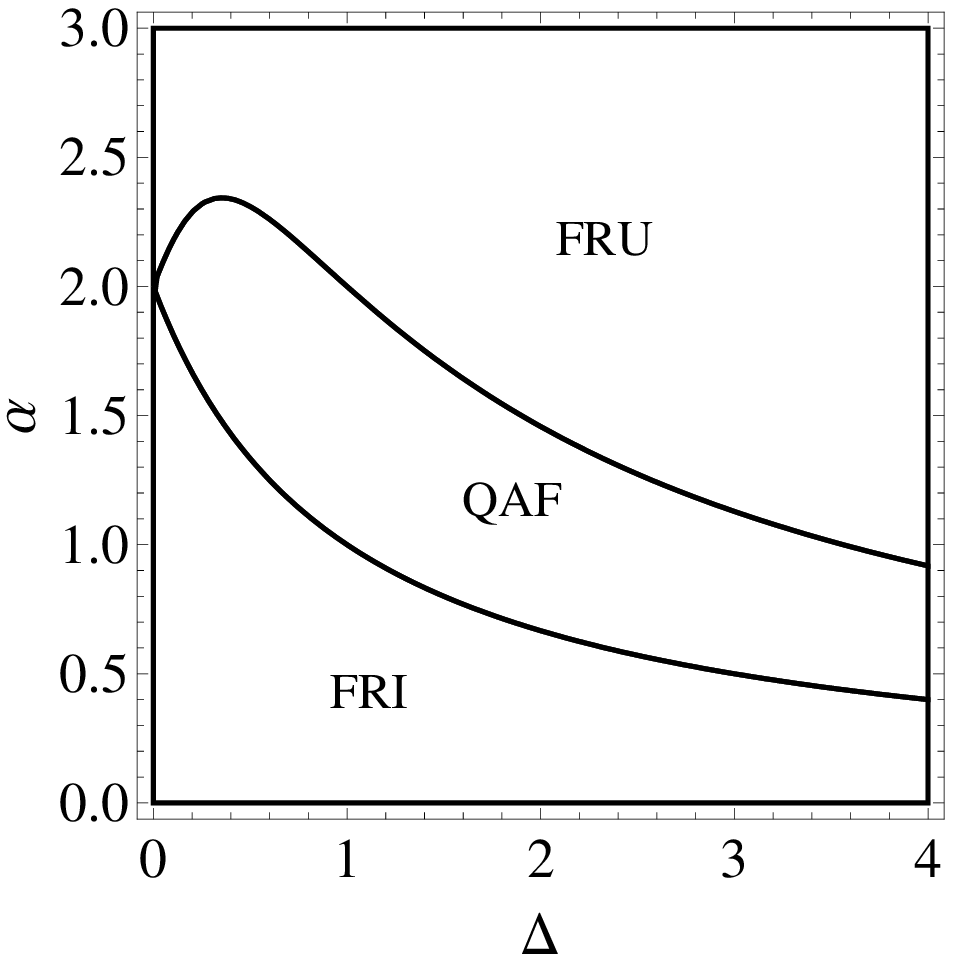} &
\includegraphics[width=5cm]{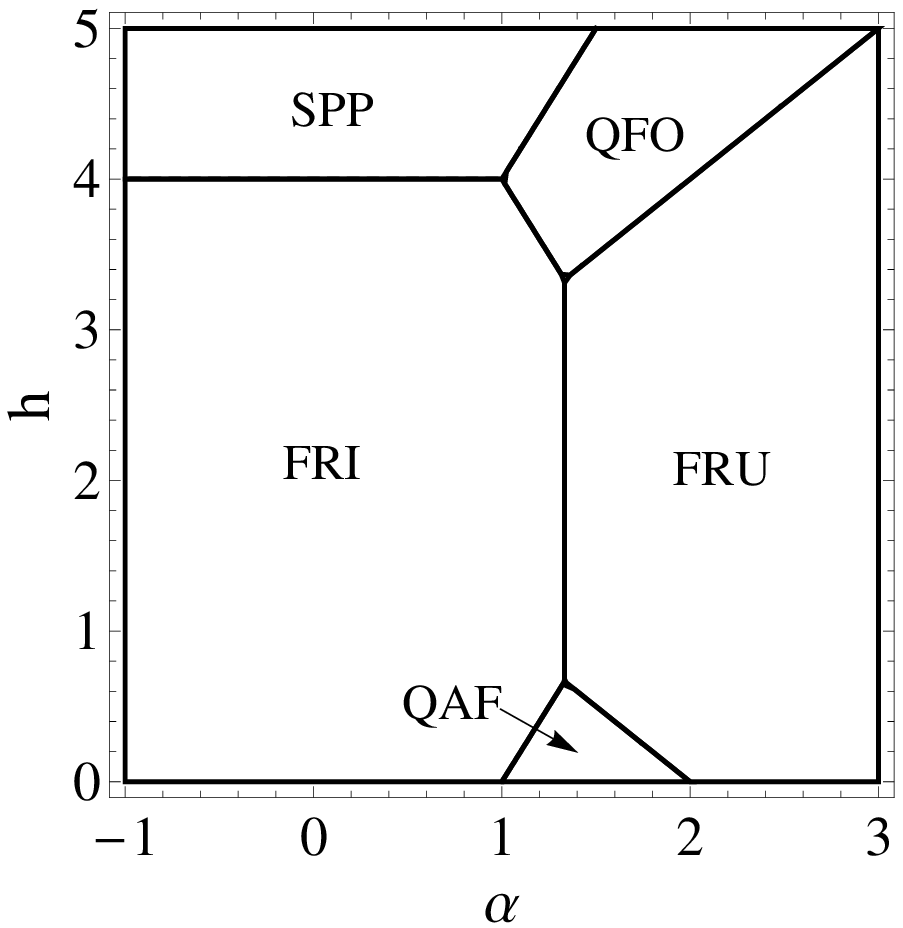}\\
\end{tabular}
\caption {\small{(a) Ground-state phase diagram in the $\Delta-\alpha$ plane in the absence of the external magnetic field $h=0$;
(b) Ground-state phase diagram in the $\alpha-h$ plane for the special value of the isotropic Heisenberg interaction $\Delta=1$.}}
\label{Phase}
\end{center}
\end{figure}

Let us establish first ground-state phase diagrams in zero and non-zero magnetic field. Only three different ground states are available in a zero magnetic field to be further referred to as the  classical ferrimagnetic phase (FRI), the quantum antiferromagnetic phase (QAF) and the frustrated phase (FRU). The respective ground states can be unambiguously characterized by the following eigenvectors, the ground-state energy per elementary block and single-site magnetizations:
\begin{itemize}
\item The classical ferrimagnetic phase:
\begin{eqnarray}
|FRI\rangle&=&\prod^{N}_{i=1}|-1 \rangle_{i} \otimes |1,1 \rangle_{a_i,b_i}, \nonumber\\
E_{FRI}&=&J-4J_1-H, \; \; \; m_I=-1, m_H=1, m=1/3,
\label{2.4a}
\end{eqnarray}
\item The quantum antiferromagnetic phase:
\begin{eqnarray}
|QAF\rangle&=& \left\{ \begin{array}{l}
\displaystyle \prod^{N}_{i=1}|+1 \rangle_{i} \otimes \frac{1}{\sqrt{2}}(|0,-1\rangle-|-1,0\rangle)_{a_i,b_i}, \\
\displaystyle \prod^{N}_{i=1}|-1 \rangle_{i} \otimes \frac{1}{\sqrt{2}}(|+1,0\rangle-|0,+1\rangle)_{a_i,b_i},
                        \end{array} \right. \nonumber \\
E_{QAF}&=&-J \Delta -2J_1, \; \; \; m_I=\pm1, m_H=\mp0.5, m=0,
\label{2.4b}
\end{eqnarray}
\item The frustrated phase:
\begin{eqnarray}
|FRU\rangle&=&\prod^{N}_{i=1}|\pm1,0  \rangle_{i} \otimes
\frac{2}{\sqrt{8+(\frac{-1+\delta}{\Delta})^2}}(|1,-1\rangle+|-1,1\rangle+\frac{1-\delta}{2\Delta}|0,0\rangle)_{a_i,b_i}, \nonumber\\
E_{FRU}&=&-\frac{J}{2}(1+\delta), \; \; \; m_I=0, m_H=0, m=0.
\label{2.4c}
\end{eqnarray}
\end{itemize}
Note that the first (second) ket vector $|\pm1,0 \rangle_{i}$ ($|\pm1,0\rangle_{a_i,b_i}$) in tensor products (\ref{2.4a})-(\ref{2.4c}) determines spin states of the Ising (Heisenberg) spins, respectively. The ground-state phase diagram for zero magnetic field is displayed in Fig.~\ref{Phase}(a) in the $\Delta-\alpha$ plane.
It can be seen from this figure that the classical ferrimagnetic spin arrangement with opposite orientation of the Ising spins with respect to the polarized Heisenberg dimers becomes the relevant ground state for sufficiently weak values of the Heisenberg coupling. The quantum antiferromagnetic phase with a more intriguing quantum entanglement of the polarized and non-magnetic states of the Heisenberg dimers becomes the ground state at moderate values of the Heisenberg interaction. Last but not least, the strong enough Heisenberg interaction is responsible for a geometric spin frustration, which supports the most peculiar frustrated ground state with a paramagnetic character of the Ising spins arising out of the quantum entanglement of the antiferromagnetic and non-magnetic states of the Heisenberg dimers. In a non-zero magnetic field, the total magnetization of the frustrated phase equals to one-third of the saturation magnetization due to a full polarization of the frustrated Ising spins into a direction of the magnetic field and besides, one may additionally find another two ground states hereafter referred to as the quantum ferromagnetic phase (QFO) and saturated paramagnetic phase (SPP):
\begin{itemize}
\item The quantum ferromagnetic phase:
\begin{eqnarray}
|QFO\rangle&=&\prod^{N}_{i=1}|+1 \rangle_{i} \otimes \frac{1}{\sqrt{2}}(|+1,0\rangle-|0,+1\rangle)_{a_i,b_i}, \nonumber\\
E_{QFO}&=&-J \Delta +2J_1-2H, \; \; \; m_I=1, m_H=1/2, m=2/3,
\label{2.1a}
\end{eqnarray}
\item The saturated paramagnetic phase:
\begin{eqnarray}
|SPP\rangle&=&\prod^{N}_{i=1}|1 \rangle_{i} \otimes |1,1 \rangle_{a_i,b_i}, \nonumber\\
E_{SPP}&=&J+4J_1-3H, \; \; \; m_I=1, m_H=1, m=1.
\label{2.1b}
\end{eqnarray}
\end{itemize}
While the saturated paramagnetic phase observable at sufficiently high magnetic fields can be characterized by a full alignment of the Ising and Heisenberg spins into the magnetic field, the quantum ferromagnetic phase is quite similar to the quantum antiferromagnetic phase except that a relative orientation of the sublattice magnetizations of the Ising and Heisenberg spins is alike. The typical ground-state phase diagram in a non-zero magnetic field is displayed in Fig.~\ref{Phase}(b) in the $\alpha-h$ plane for the illustrative case of the isotropic Heisenberg interaction $\Delta = 1$. It is worth noticing that the total magnetization equals zero for the quantum antiferromagnetic phase, one-third of the saturation magnetization for the classical ferrimagnetic phase, either zero or one-third of the saturation value for the frustrated phase depending on whether the magnetic field is zero or not, and two-thirds of the saturation magnetization for the quantum ferromagnetic phase.

\begin{figure}[!h]
\begin{center}
\begin{tabular}{cccc}
{\small (a)}& {\small (b)}\\
\includegraphics[width=6cm]{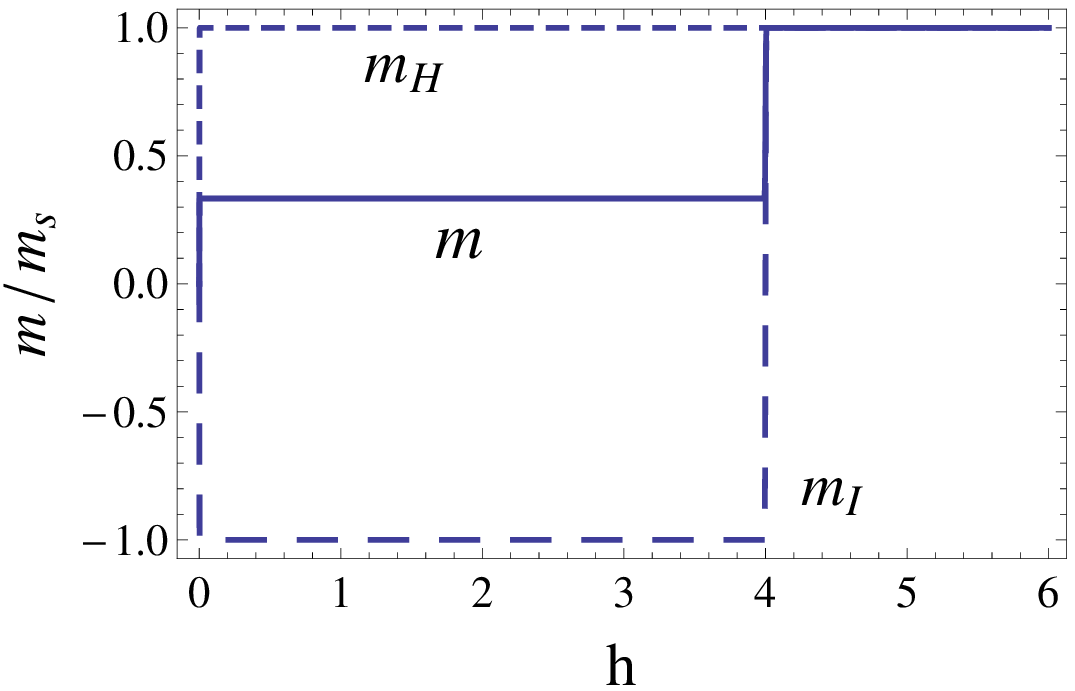} &
\includegraphics[width=6cm]{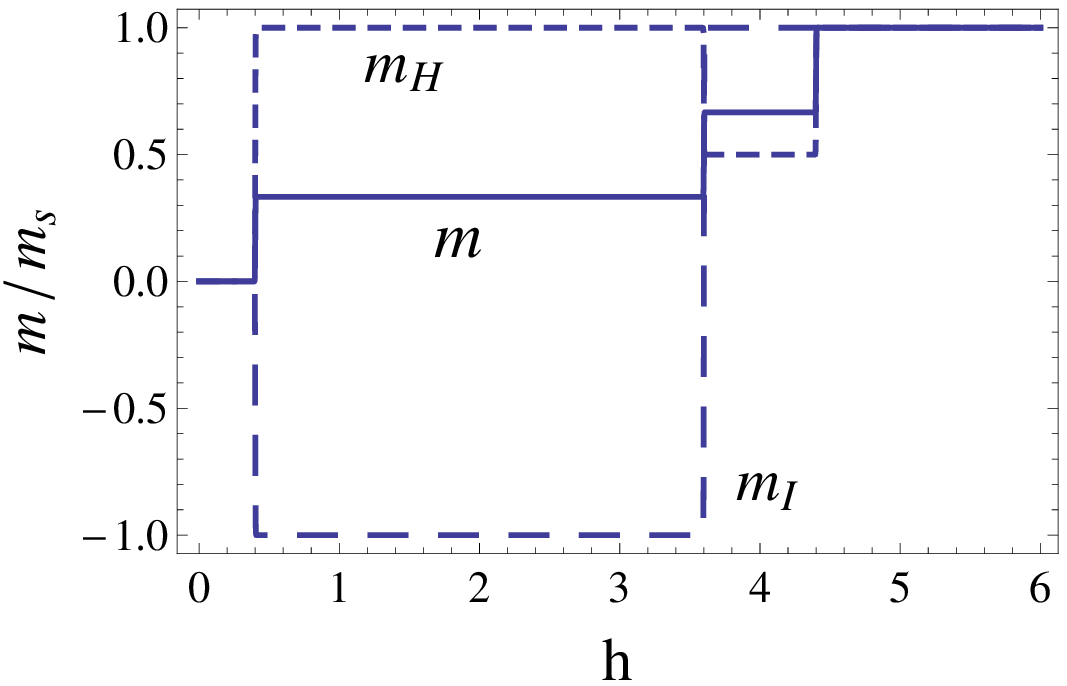}
\nonumber \\
{\small (c)}& {\small (d)}\\
\includegraphics[width=5.9cm]{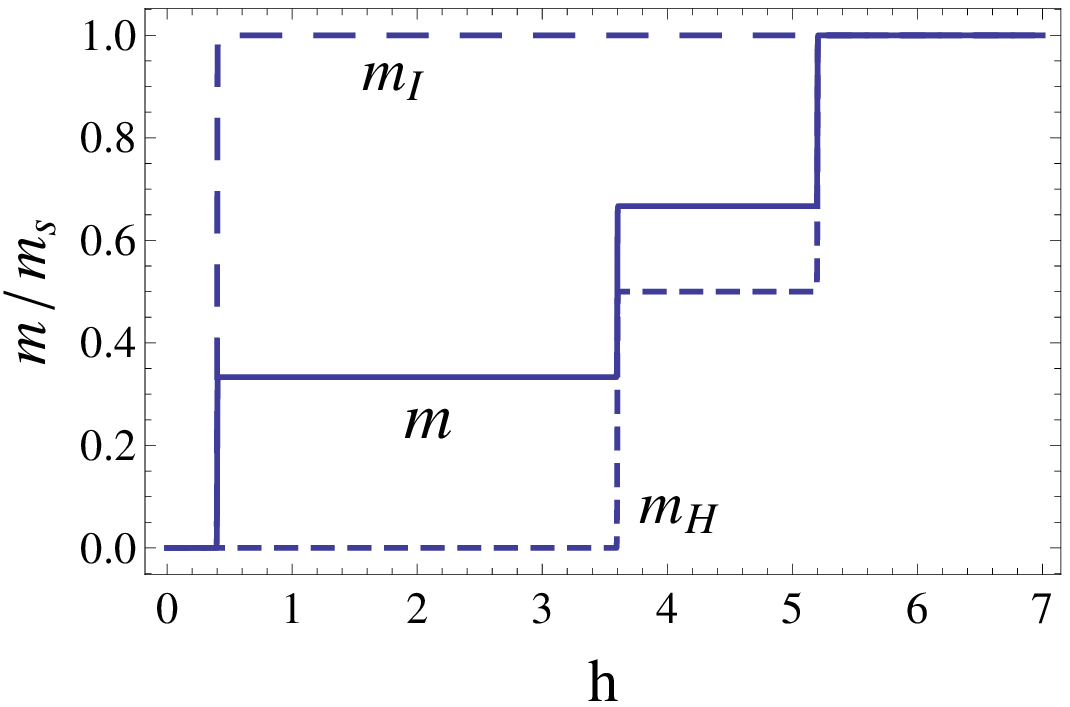} &
\includegraphics[width=5.9cm]{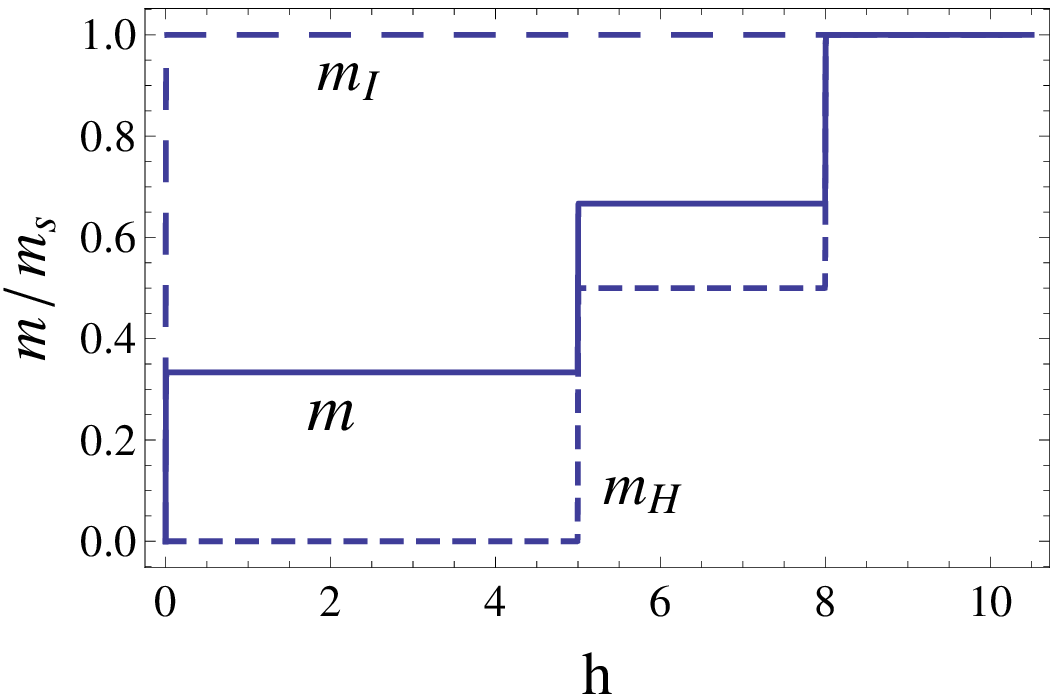}
\end{tabular}
\caption { \small{The magnetic-field dependence of the total and sublattice magnetizations at low enough temperature ($t=0.001$) for the fixed value of the anisotropy
parameter $\Delta=1$ and a few different values of the interaction ratio $\alpha$ (a) $\alpha=-2$ (b)  $\alpha=1.2$, (c) $\alpha=1.6$, (d) $\alpha=3$.}}
\label{Mag2}
\end{center}
\end{figure}

The ground-state phase diagram in a non-zero magnetic field implies various magnetization scenarios depending on a relative strength of the Heisenberg and Ising interactions. To verify this issue, we have plotted in Fig.~\ref{Mag2} typical low-temperature magnetization curves supporting four different types of magnetization scenarios: (a) FRI-SPP, (b) QAF-FRI-QFO-SPP, (c) QAF-FRU-QFO-SPP, (d) FRU-QFO-SPP. Although the magnetization of the spin-1 Ising-Heisenberg diamond chain varies continuously with the magnetic field at any finite temperature, the magnetization shows a rather steep increase with increasing magnetic field at low enough temperatures that indeed resembles zero-temperature magnetization jumps closely connected to the field-induced transitions between different ground states. It could be concluded that the the spin-1 Ising-Heisenberg diamond chain may display a rather diverse magnetization process including either one, two or three intermediate magnetization plateaus. If the frustration parameter is sufficiently small $\alpha<1$, then, one observes just one intermediate plateau at one-third of the saturation magnetization related to the classical ferrimagnetic phase (Fig.~\ref{Mag2}a). The most spectacular magnetization curve with three different intermediate plateaus at zero, one-third and two-thirds of the saturation magnetization can be detected for moderate values of the frustration parameter $1<\alpha<2$, which supports in general a presence of the quantum antiferromagnetic phase with zero magnetization at low enough magnetic fields and the quantum ferromagnetic phase with the total magnetization equal to two-thirds of the saturation value for higher magnetic fields slightly below the saturation field (see Fig.~\ref{Mag2}b-c). In addition, the one-third magnetization plateau appears at moderate values of the magnetic field, which is either connected to the classical ferrimagnetic phase for relatively weaker values of the frustration parameter $1<\alpha<4/3$ (Fig.~\ref{Mag2}b) or the frustrated phase for relatively stronger values of the frustration parameter $4/3<\alpha<2$ (Fig.~\ref{Mag2}c). Last but not least, the intermediate plateau at zero magnetization totally disappears from the magnetization curve whenever the frustration parameter $\alpha>2$ is strong enough to ensure the frustrated phase in zero magnetic field (Fig.~\ref{Mag2}d).

\begin{figure}[!h]
\begin{center}
\begin{tabular}{ccc}
{\small (a)}& {\small (b)}& {\small (c)}\\
\includegraphics[width=5.7cm]{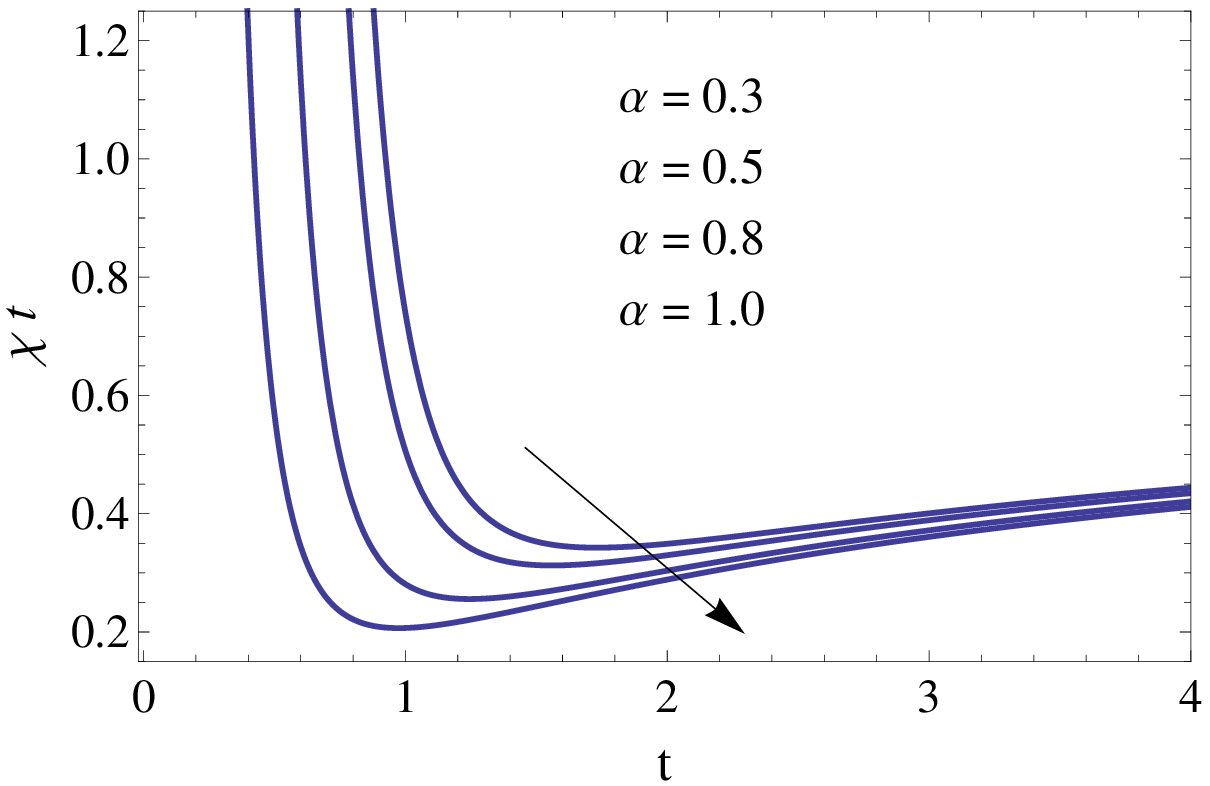} &
\includegraphics[width=5.8cm]{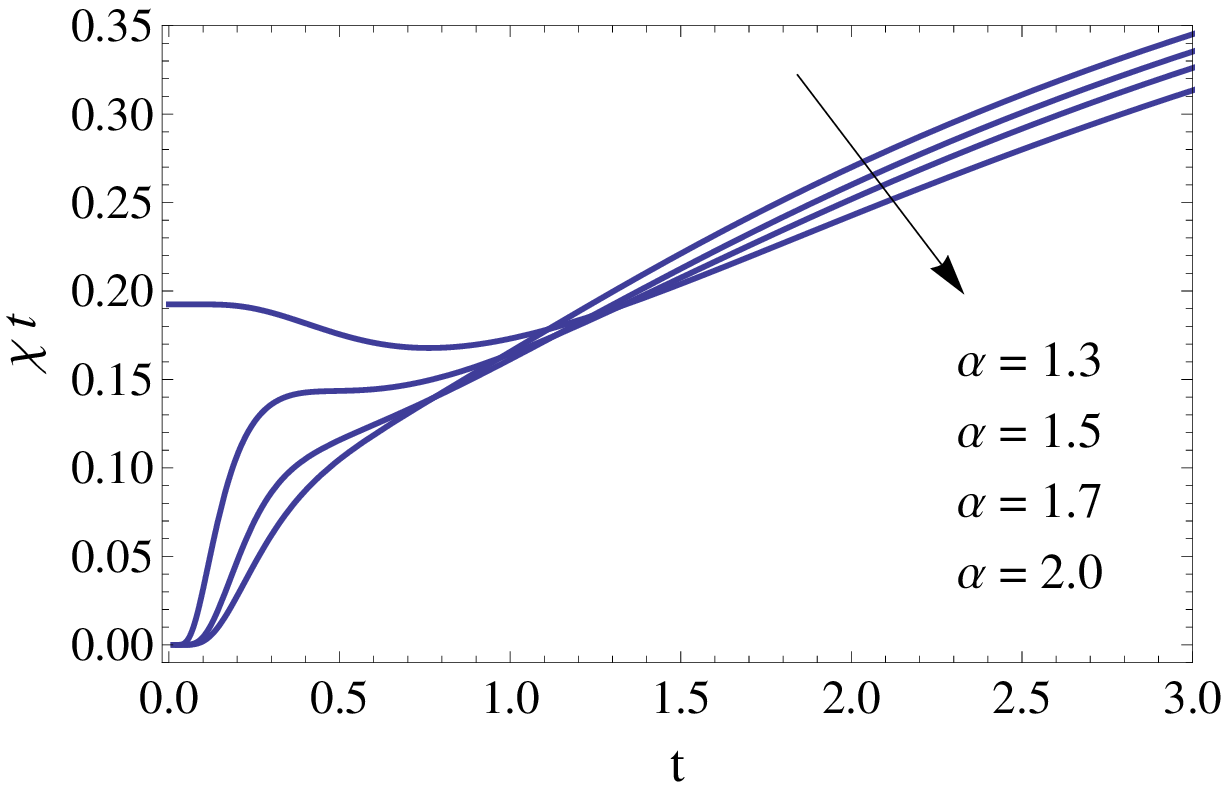} &
\includegraphics[width=5.7cm]{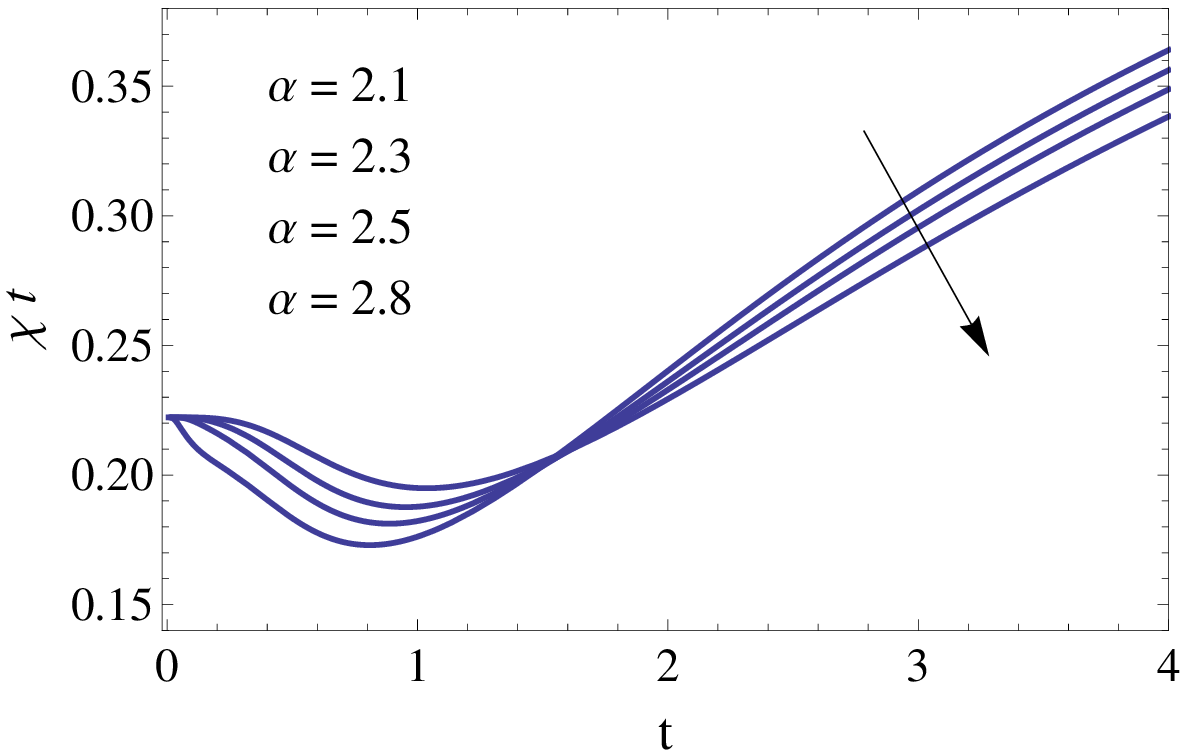}
\end{tabular}
\caption {\small{Typical thermal variations of the zero-field susceptibility times temperature product for the fixed value of the anisotropy parameter $\Delta=1$ and a few different values of the
interaction ratio $\alpha$: (a) $\alpha=0.3, 0.5, 0.8, 1.0$, (b) $\alpha=1.3, 1.5, 1.7, 2.0$, (c) $\alpha=2.1, 2.3, 2.5, 2.8$.}}
\label{Sus}
\end{center}
\end{figure}

Finally, let us turn our attention to typical temperature dependences of the zero-field susceptibility times temperature product as depicted in Fig.~\ref{Sus}. The interaction ratio $\alpha$ in three different panels of Fig.~\ref{Sus} has been chosen so as to support one of three available zero-field ground states: the classical ferrimagnetic phase for $\alpha<1$, the quantum antiferromagnetic phase for $1<\alpha<2$ and the frustrated phase $\alpha>2$, respectively. Fig.~\ref{Sus}a illustrates thermal dependences of the susceptibility times temperature product for the classical ferrimagnetic phase with a low-temperature divergence and round minimum at moderate temperatures, which are quite typical for one-dimensional ferrimagnets. Contrary to this, the susceptibility times temperature product tends towards zero when approaching zero temperature for moderate values of the interaction ratio $\alpha$ supporting the quantum antiferromagnetic phase (Fig.~\ref{Sus}b). The last possible scenario is illustrated in Fig.~\ref{Sus}c for the frustrated phase supported by greater values of the interaction ratio $\alpha$, which shows an outstanding non-monotonous thermal dependence of the susceptibility times temperature product approaching in the zero-temperature limit a finite asymptotic value on account of the paramagnetic (frustrated) character of the Ising spins.

\section{Conclusion}
\label{conclusion}

In the present work, we have examined the ground state and magnetic properties of the exactly solvable spin-$1$ Ising-Heisenberg diamond chain in a magnetic field. The exact solution of the
model under investigation has been obtained within the framework of the transfer-matrix method. Our particular attention has been focused on the ground-state phase diagrams in zero and non-zero magnetic fields, typical magnetic-field dependences of the total and sublattice magnetizations, as well as, typical temperature variations of the zero-field susceptibility times temperature product.

It has been evidenced that the spin-$1$ Ising-Heisenberg diamond chain exhibits an outstanding diversity of magnetization curves, which may include either one, two or three intermediate magnetization plateaus at zero, one-third and two-thirds of the saturation magnetization. In this regard, we have verified that all intermediate magnetization plateaus admissible by the OYA rule for the symmetric spin-$1$ diamond chain without translationally broken symmetry can be indeed present in a magnetization process.

Finally, it is worthwhile to remark that the rigorous procedure elaborated in the present work can be rather straightforwardly extended to a more general spin-$1$ Ising-Heisenberg diamond chain accounting for the asymmetric Ising interactions along diamond sides, further-neighbour interaction between the Ising spins, the single-ion anisotropy and/or biquadratic interaction. Our preliminary results for a more general spin-$1$ Ising-Heisenberg diamond chain indeed imply a greater diversity of its magnetic features including greater number of intermediate magnetization plateaus and magnetization scenarios \cite{hov14}.

\ack{J S acknowledges financial support provided by a grant from The
Ministry of Education, Science, Research, and Sport of the Slovak
Republic under Contract No. VEGA 1/0234/12 and by grants from the
Slovak Research and Development Agency under Contracts No.
APVV-0132-11 and No. APVV-0097-12. N S acknowledges partial
financial support by the MC-IRSES No. 612707 (DIONICOS) under
FP7-PEOPLE-2013 and research project No. SCS 13-1C137 grants. V H
would like to acknowledge support from the NFSAT, the ''Young
Scientists Support Program'' (YSSP), Youth Foundation of Armenia
(AYF), and CRDF Global under grant YSSP-13-02.}

\end{document}